\documentclass{PoS}

\title{Estimation of the exposure for the air shower detection mode of EUSO-SPB1}

\ShortTitle{Estimation of the exposure for the air shower detection mode of EUSO-SPB1}
\usepackage{wrapfig}
\usepackage{upgreek}

\author{\speaker{K.~Shinozaki}, M.~Bertaina, F.~Bisconti\thanks{Affiliated to
    INFN}, F.~Fenu, S.~Ferrarese, S.~Monte\thanks{Affiliated to University of
    Turin}\\
  University of Turin and/or INFN Torino, Turin, Italy
  
  E-mail: \email{kenji.shinozaki@to.infn.it}}

\author{A.~Anzalone, A.~Bruno\\
  INAF\,--\,IASF Parlermo, Palermo, Italy}

\author{{S.~Briz}\\
  University Carlos III de Madrid, Madrid, Spain}

\author{{A.~Diaz}\\
  IRAP, University of Toulouse, Toulouse, France}

\author{{J.~Eser, L.~Wiencke}\\
  Colorad School of Mines, Golden, USA}

\author{{A.~Olinto}\\
  University of Chicago, Chicago, USA}

\author{{M.~Vrabel}\\
  Technical University of Kosice, Kosice, Slovakia}

\author{{for the JEM-EUSO collaboration}\footnote{for collaboration list see PoS(ICRC2019)~1177}}

\ShortTitle{Estimation of the exposure for the air shower detection mode of EUSO-SPB1}

\abstract{EUSO-SPB1 was a balloon-borne pathfinder mission of the
  JEM-EUSO~(Joint Experiment Missions for the Extreme Universe Space 
  Observatory) program. A 12-day long flight started from New Zealand
  on April 25~th, 2017 on-board the NASA's Super Pressure Balloon. With capability of
  detecting EeV energy air showers, the data acquisition was performed using a
  1~m$^2$ two-Fresnel-lens UV-sensitive telescope with fast readout
  electronics in the air shower detection mode over $\sim$~30~hours at
  $\sim$16\,--\,30~km above
  South Pacific. Using a variety of approaches, we searched for air shower
  events. Up to now, no air shower events have been identified. 
  The effective exposure, regarding the role of the clouds in particular,  
  was estimated based on the air shower and detector simulations together
  with a numerical weather forecast model. Compared with the case assuming
  the fully clear atmosphere conditions, more than $\sim$60\% of showers
  are detectable regardless the presence of the clouds. The studies in the
  present work will be applied in the follow-up pathfinders and in the future
  full-scale missions in the JEM-EUSO program.
}
\FullConference{36th International Cosmic Ray Conference -ICRC2019-\\
		July 24th - August 1st, 2019\\
		Madison, WI, U.S.A.}

\begin{document}

\section{Introduction} 
\label{intro} 
The origin of the ultra-high energy cosmic rays~(UHECRs; referred to as
$\sim$EeV energies and higher) is still an open question in the today's
astrophysics~\cite{Dawson17}. In particular, above $\sim$100~EeV, the UHECR
fluxes are on the an order of 1~per~km$^2$ in a millennium or even fewer with a
steepening spectral index which is interpreted as the well-known
Greisen-Zatsepin-Kuz'min mechanism during the propagation of UHECRs or/and
acceleration limit of UHECRs at their sources. To further investigate the
nature of UHECRs, larger exposures are essential. 

The JEM-EUSO~(Joint Experiment Missions for Extreme Universe Space
Observatory) program~\cite{Bertaina19} aims at space-based UHECR observations
from satellite orbits. Using an ultra-wide field-of-view~(FOV) UV telescope,
the fluorescence technique enhances instantaneous apertures to efficiently
increase the exposure. In our framework, two full-scale missions 
are promoted as K-EUSO~(KLYPVE-EUSO)~\cite{Klimov19} and POEMMA~(Probe Of 
Extreme Multi-Messenger Astrophysics)~\cite{Olinto19}. The former 
will be based on the International Space Station~(ISS).
To test the key technologies developed for JEM-EUSO, we had conducted
pathfinders, a total of 130~hours of operation of EUSO-TA~\cite{Bisconti19} at
the Telescope Array~(TA)~\cite{Tokuno12} site in Utah, USA and the 
EUSO-Balloon~\cite{Abdellaoui19} on the one night stratospheric CNES~(French
Space Agency) balloon flight in Timmins, Canada.

In space-based UHECRs observations, several factors are needed to be taken 
into account~\cite{Adams13}. In particular, the trigger conditions should be 
optimized to fit the telemetry limit. Air shower signals originate in the 
atmosphere behind the airglow background light that is persistently emitted in 
the mesosphere altitude around 90~km~above sea level. The intensity of such 
diffuse light determines threshold energy of detectable UHECR events. The 
presence of the clouds in the FOV partly reduced the instantaneous apertures 
in terms of area and solid angle acceptance. 

In April 2017, we conducted the EUSO-SPB1 mission to operate the first 
fluorescence telescope flown to sub-orbital space with the capability of air 
shower detection. In the present work, we evaluated the effective exposure for 
EUSO-SPB1 mainly focusing on the role of the clouds. We will remark the
relevant issues in the future space-based UHECR observations. 

\section{EUSO-SPB1 mission} 

The EUSO-SPB1 telescope~\cite{Eser19} consisted of a 1m$^2$ aperture
two-Fresnel-lens optics and a photo-detector module~(PDM)~\cite{Casolino15}
composed of 36~multi-anode photomultiplier tubes~(MAPMTs) covered with
BG3~UV-band-pass 
filters. The total of 2304~pixels cover an $\sim$11$^\circ$~FOV  with 
$\sim$0.2$^\circ$ spatial pitch and small non-efficient parts among MAPMTs.
In all the pixels, photon counting is performed every 2.5~$\upmu$s. This time
duration is called a gate time unit~(GTU).

Compared with other JEM-EUSO pathfinders, the major characteristics of
EUSO-SPB1 was an autonomous trigger to detect air showers in the EeV
regime~\cite{Bayer17}. Once a trigger is issued, the counts on all the pixels
were saved for 128~GTUs. Prior to the flight, a field test of EUSO-SPB1 was
carried out at the TA site~\cite{Tokuno12} to verify and to
quantify the trigger and other key functions using a movable laser device and
the TA's Central Laser Facility. By detecting signals from the scattered
light of laser shots, we evaluated the performance of EUSO-SPB1~\cite{Eser19}.

\begin{figure*}[t]
  \centering
  \begin{tabular}{cc}
    \centering
    \parbox{0.5\textwidth}{      
      \includegraphics[width=0.5\textwidth]{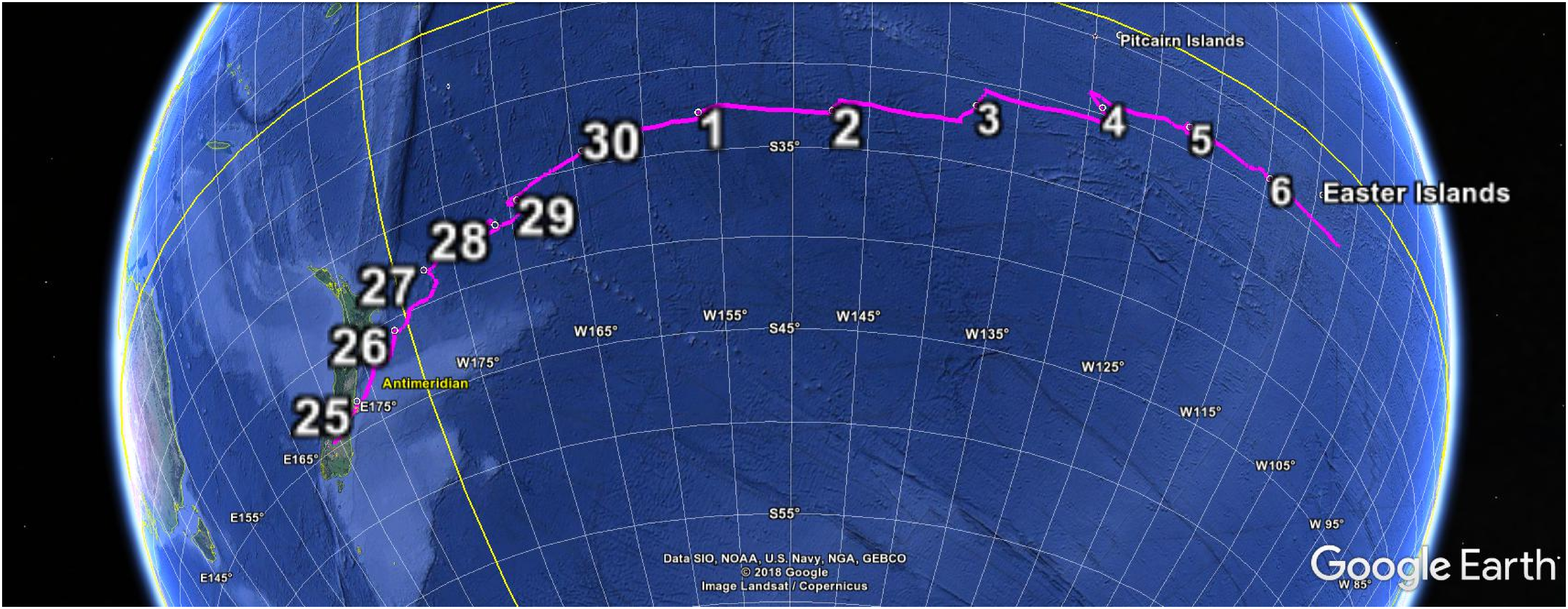} 
      \caption{EUSO-SPB1 flight track. The 
        positions at 12:00~UTC are labeled by the date.}
      \label{fig1}}&
    \parbox{0.4\textwidth}{          
      \includegraphics[width=0.4\textwidth]{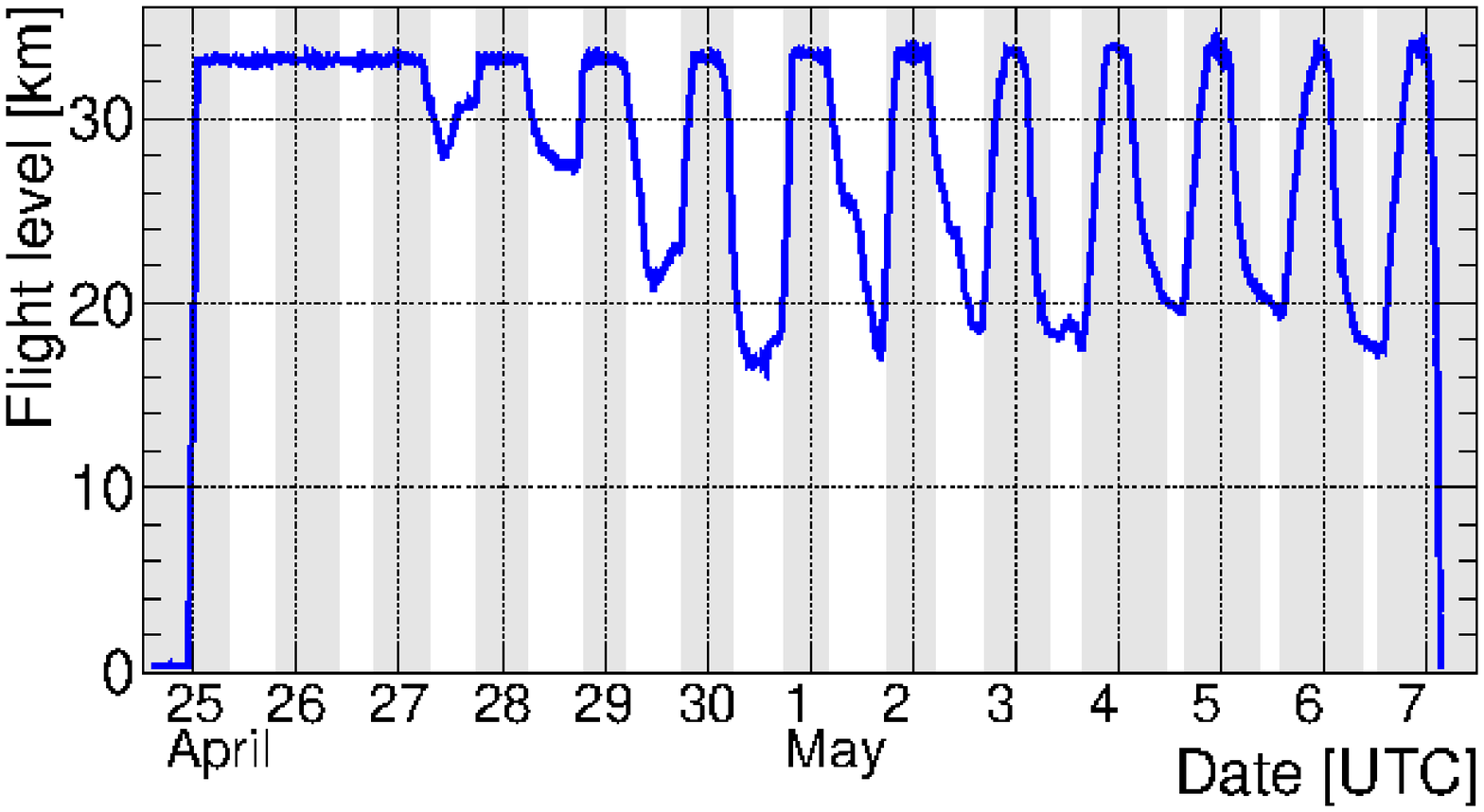} 
      \caption{EUSO-SPB1 flight level. Bright regions
        indicate hours with any DAQ made.}   \label{dath} }
  \end{tabular}
\end{figure*}

\begin{figure*}[t] 
  \centering 
  \includegraphics[width=0.66667\textwidth]{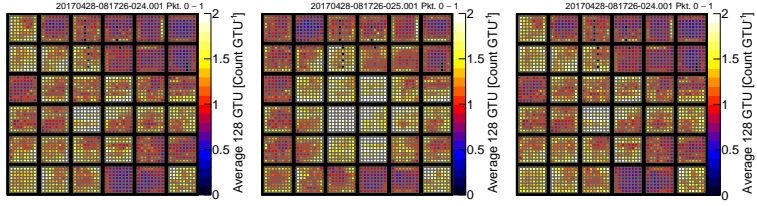} 
  \caption{Example of the background images on the FOV measured by the
    EUSO-SPB1 telescope. These images are the averaged 10~adjacent triggers.
    Each panel corresponds to $\sim$1~min time interval.} 
  \label{pass} 
\end{figure*}

Figure~\ref{fig1} shows the EUSO-SPB1 track on the Google Earth image.
Positions as of 12:00~UTC are labeled with the day of the month. EUSO-SPB1
resided around the Anti-meridian. This time corresponding to around the local
midnight~(hereafter nights mean the local nights of UTC dates).

Figure~\ref{dath} displays the variation of EUSO-SPB1 flight level. Bright
regions indicate hours in which any data acquisition~(DAQ) was made by the
EUSO-SPB1 telescope.

EUSO-SPB1 onboard NASA's Super Pressure Balloon~(SPB) was released from the
Wanaka airport, New Zealand on April~24th, 2017 at 22:50~UTC, around the noon
of April~25th local time. EUSO-SPB1 first reached the nominal flight level of
$\sim$33~km in 2~hours. Since April 27th, the flight level tended to drop due
to the unexpected gas leakage on the balloon envelope. While the efforts for
recovering EUSO-SPB1 on the continent, the flight level finally became
$\sim$17~km. The flight lasted $\sim$12.4~days over $\sim$7000~km until we had
to abandon the payload off Easter Island.
%at Latitude of 29.38$^\circ$S and Longitude of106.50$^\circ$W, 

Since 08:31~UTC of April 25th, we operated the air shower detection mode every
night for $\sim$30~hours in total. The operable time decreased night by night
due to the waxing Moon in the sky. In the end, a $\sim$27~hours of the data, in part
reduced to a 25~GTU length were transmitted to the ground before EUSO-SPB1 sunk.
Excluding commissioning phase, $\sim$21.7~hours of the data were acquired when
all MAMPTs were operational. The EUSO-SPB1 instrument had been operational till
the end. We confirmed the functionality of the trigger from trigger rates and
background levels~\cite{Bayer17}. Using a variety of approaches from visual
inspection to machine learning, we search the triggered events for air
shower event. Up to now, no candidates have been found~\cite{Diaz19,Vrabel19}.

Figure~\ref{pass} shows selected images of background in the FOV of the 
EUSO-SPB1 telescope. These images are averaged over the 10~adjacent triggers, 
a time span of 3.2~ms. Each panel corresponds to $\sim$1~min time 
interval~\cite{Shinozaki19}. This figure is also intended to illustrate the 
deployment of MAMPTs in the PDM. The color scales denote the background level 
in units of counts~per GTU. 

In these images, the cloud pattern and its relative motion with EUSO-SPB1 was 
also recognized. Over its flight, EUSO-SPB1 oversaw various weather
situations. To monitor the presence of clouds in the FOV, an infrared camera
was operated in limited times along with the EUSO-SBP1 telescope and also in
the daytime. Preliminary results based on a possible method~\cite{Anzalone19}
that could be applied to the data are presented in Reference~\cite{Bruno19}. 

\begin{figure*}[!t] 
  \centering 
  \includegraphics[width=0.85\textwidth]{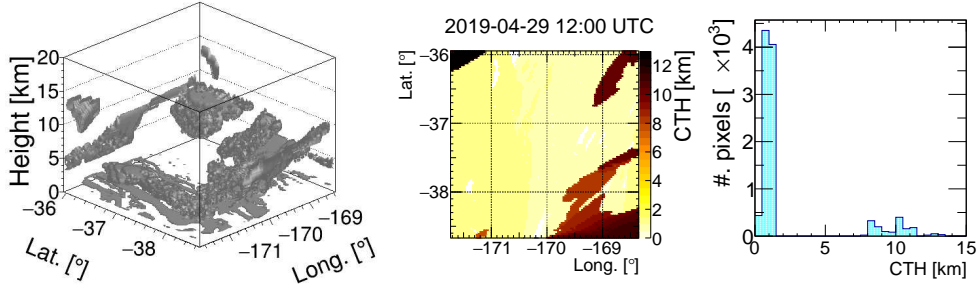} 
  \caption{A WRF output for 12:00~UTC of April 29th. Left:~visualization of
    grid-boxes with a non-zero cloud fraction. Middle:~CTH map around the
    EUSO-SPB1 position. Right:~CTH distribution in this area.} 
  \label{wrfsample} % Give a unique label 
\end{figure*} 

\section{Cloud-top height estimations and air shower simulations} 

The role of clouds is important factor to the performance of the air shower 
detections for EUSO-SPB1 and for future pathfinders and full-scale missions.  
In the present work, we utilized the WRF~(Weather Research and 
Forecasting)~model~\cite{NCAR} to estimate the cloud-top height~(CTH).
WRF is a mesoscale meteorological model developed by the National Center for
Atmospheric Research. It can reach up to a 1 km horizontal spatial resolution.
WRF provides the atmospheric quantities for each pixel over all vertical levels
on the model grid. Further details for this specific study is given in
Reference~\cite{Monte19}. The key result from WRF was presented in
Reference~\cite{Shinozaki19}. Regional weather forecasting models
in general simulate the atmospheric quantities staring with an input of a
global circulation model. We employed ECMWF~(European Centre for Medium-range
Weather Forecasts)~data~\cite{ecmwf} for such inputs. CTH retrieval was based
on the cloud fraction which expresses the cloudiness in a model grid-box.
Outputs are given in 10-minute intervals for grid-boxes of a $\sim$3 km pixel
over 49~pressure levels below $\sim$~30~km height. We defined CTH by the
highest grid-box with a cloud fraction greater than 0.2 on horizontal pixel. 

Figure~\ref{wrfsample} displays a WRF output for 12:00~UTC of April~29th. The  
left panel visualizes grid-boxes with a non-zero cloud fraction seen from  
southwest. The middle and right show the CTH map and distribution in a 
$\sim$300~km$\times$300~km~area around the EUSO-SPB1 position, respectively. 

\begin{figure}[t]
  \centering 
  \includegraphics[width=0.8\textwidth,height=3cm]{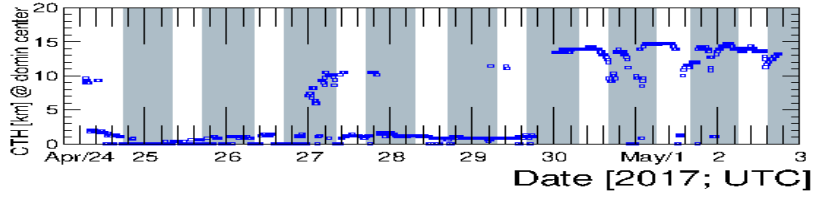} 
  \vspace*{-1em}
  \caption{CTH variation between April~24th and May~2nd. Bright regions are
    night time at EUSO-SPB1.}
  \label{cthdate} % Give a unique label 
\end{figure}

Figure~\ref{cthdate} displays the CTH variation from April~24th through
May~2nd. Bright regions denote the night time between twilights at the
EUSO-SPB1 position. 

The WRF results show that the CTH variation was moderately small and was as 
low as a few kilometers until April~29th. In the following nights, CTH reach  
the height of tropopause. The CTH variation are used for studying the cloud 
impact on the air shower detection with the simulations. 

So far, we have estimated how many air shower events could be detected during
the EUSO-SPB1 flight. Using the air shower and detector simulations with `EUSO
OffLine'~\cite{Paul17}, it turned out to be 1.2$\pm$0.4~event for 21.7~h
assuming the clear atmospheres~\cite{Eser19}. These simulations take into
account the calibration of the EUSO-SPB1 telescope in the laboratory. The peak
energy of the triggered events has been found consistent with the laser
detection results in the field test. 

In the present work, we investigate the role of the cloud in the air shower 
detection using the ESAF~(EUSO Simulation and Analysis
Framework)~software~\cite{Berat10}. For the air shower detection from the
above, 
CTH and the optical depth of the cloud are key factors for the detection 
efficiency. For a quantification of the cloud impact during the EUSO-SPB1  
flight, we assumed the CTH variation as shown in Figure~\ref{cthdate}. In the
former work~\cite{Adams13}, we defined the 
parameter called `cloud efficiency' $\kappa_{\rm C}$ as the ratio in apertures 
of air shower detections, namely equivalent to the ratio in event rates 
estimated by assuming the modeled cloudy conditions to the those estimated 
under the clear atmosphere conditions. 

To determine this value for EUSO-SPB1, we first simulated the air showers
with the case there is no clouds in the FOV by assuming a omni-directional
isotropic UHECR flux given in 
Reference~\cite{Fenu17}. The background level was 
taken from the average of the real data~\cite{Bayer17,Shinozaki19}. 
The event rates for three typical flight levels of 17~km, 21~km, and 28~km 
were estimated. The overall rate was estimated by weighting the durations of 
DAQ in different levels. To correct uncertainty in the energy scale due to the 
models used in ESAF and OffLine, we re-scaled to fit the above mentioned event 
rate obtained by OffLine. 

\begin{figure*}[t]
  \centering
  \begin{tabular}{cc}
    \parbox{0.45\textwidth}{      
      \includegraphics[width=0.45\textwidth]{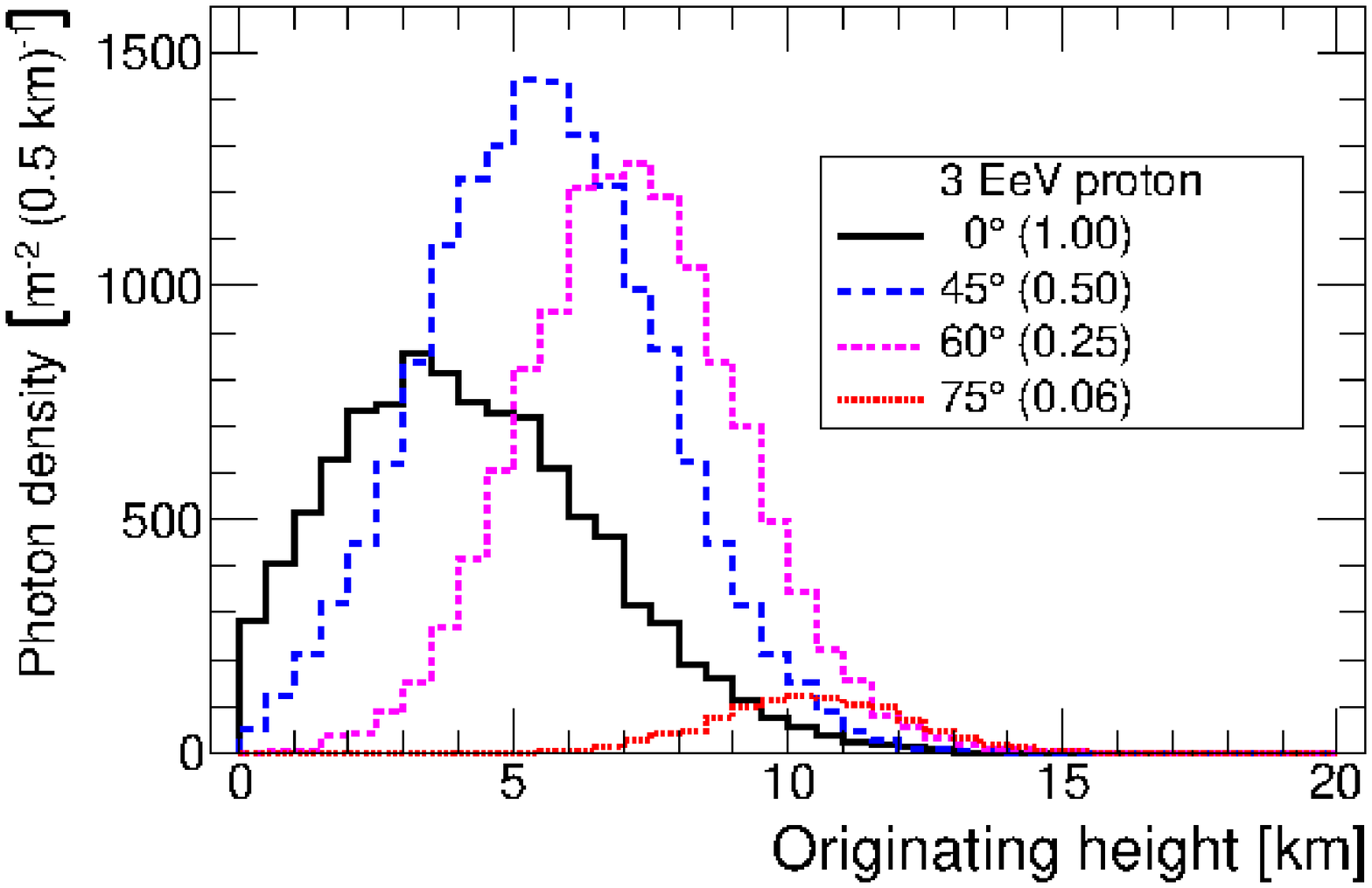}
      \caption{Fluorescence photon density at a 28~km~height as a function of
        the originating height. Air showers from a 3~EeV proton were simulated
        towards the sub-balloon point.}
      \label{hflux}}&
    \parbox{0.45\textwidth}{          
      \includegraphics[width=0.45\textwidth,height=4.cm]{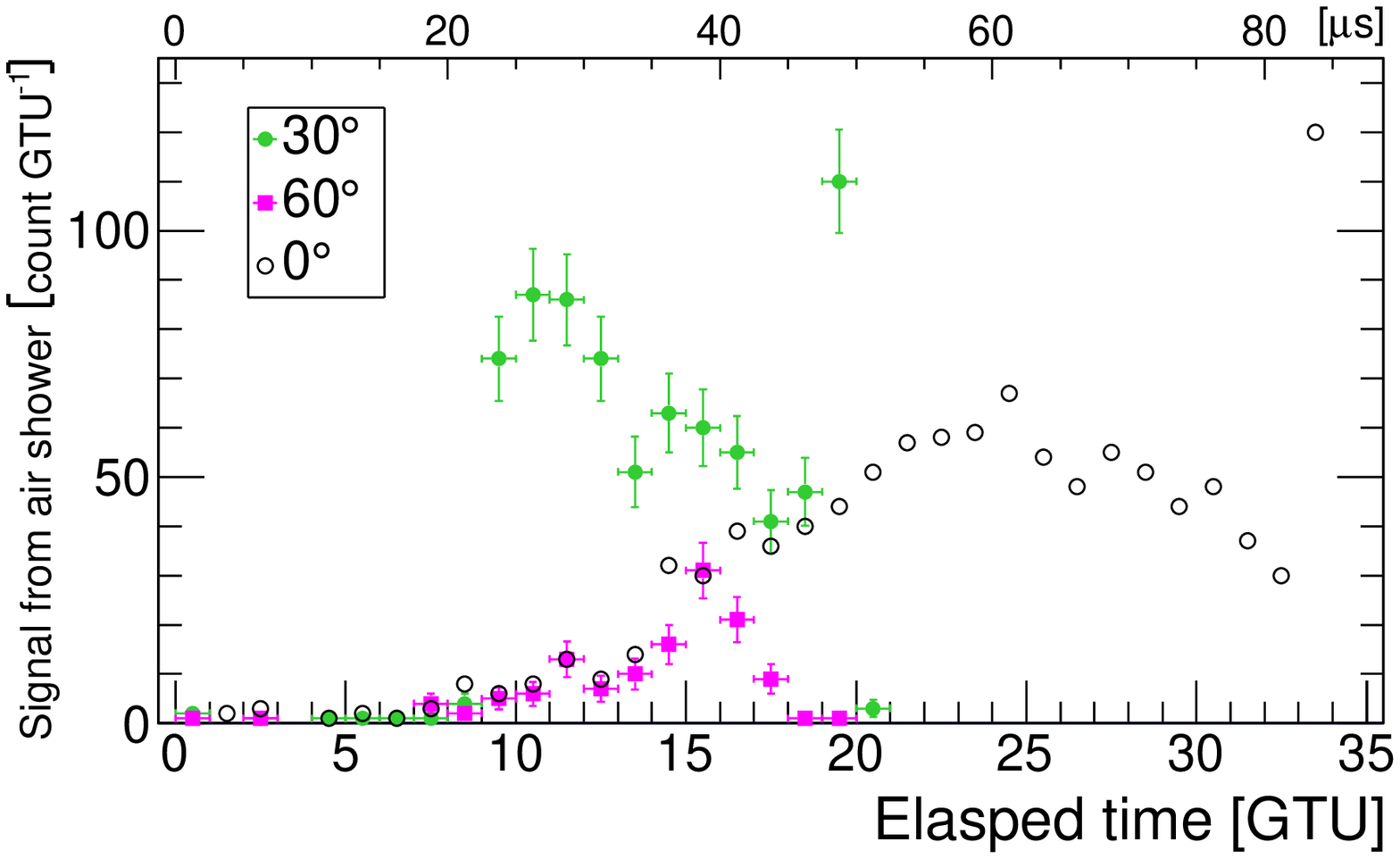} 
      \caption{Examples of simulated air shower signals on the PDM with the 
        same conditions as in Figure~\ref{hflux} but for three different 
        zenith angles. The horizontal axis is elapsed time since the first 
        signal detection.} 
  \label{signal}}
  \end{tabular}
\end{figure*}

Figure~\ref{hflux} displays fluorescence photon density at a EUSO-SPB1 level
of 28~km as a function of the originating height. In the clear atmospheres, air
showers induced by a 3~EeV proton towards the sub-balloon point on sea level.
Different incident zenith angles~$\vartheta=0^\circ$, $45^\circ$, $60^\circ$,
and $75^\circ$, were simulated. The effect of Rayleigh scattering by molecules
in the air and absorption by ozone is included. This figure is intended to
illustrate from which height how many photons reach the EUSO-SPB1 telescope
regardless the FOV covered by the PDM. 

Within a limited energy range, the heights where air showers develop depend on 
the incident zenith angles, i.e., the larger zenith angles, air showers
develop the closer to the observation level. On the other hand, as seen for
the case of $\vartheta=75^\circ$, the location of the shower development can
displace far away from the FOV. For a lower flight level of 17~km for example,
the characteristics of this figure are modified by other factors. Air showers
may develop even a few kilometers away and lower density of the air allows more
photons to reach EUSO-SPB1. The vertical 
depth corresponds to $\sim$90~g~cm$^{-2}$. The mean free path of UHECRs 
in EeV energies is $\sim$45\,--\,60~g~cm$^{-2}$ in the
air~\cite{Abbassi15,Abreu12}. Thus, most of UHECRs initiate air showers above
this level.

Figure~\ref{signal} display examples of simulated air shower signals on the PDM
with the same conditions as in Figure~\ref{hflux} but for zenith angles of
$30^\circ$ and $60^\circ$ cases, and $0^\circ$ case  for comparison. 

An unlikely $\vartheta=0^\circ$ example represents a case that the whole
shower development takes place within the FOV. The largest excess
is a result of Cherenkov light back-scattered from the sea. In
this case, stationary but variable signals are seen on the PDM. In other
cases, a light spot moves on the background. The time profile of the detected
signals largely depends on which part of showers is observable within a
limited FOV of EUSO-SPB1. Among these examples, only the 60$^\circ$ case does
not trigger. Typically, large zenith angles are more advantageous if the maximum of the shower development takes place near or in
the FOV, 

In the same way, simulating air showers in the cloudy conditions were made by 
assuming one-layer cloud over the whole FOV at the sampled CTH from WRF
outputs. We conservatively assumed the optical depth of cloud to be 1. Thus, 
all the photons originate below the CTH are attenuated by $\exp(-1)$ 
before reaching EUSO-SPB1 in addition to the effect suffered from the clear 
atmospheres. If `not very old' air showers reach the cloud level in the FOV, 
more intense back-scattered Cherenkov light may be seen compared with those 
from sea as is seen in Figure~\ref{signal}. 

\begin{figure*}[t]
  \centering
  \begin{tabular}{cc}
    \parbox{0.45\textwidth}{      
  \includegraphics[width=0.45\textwidth]{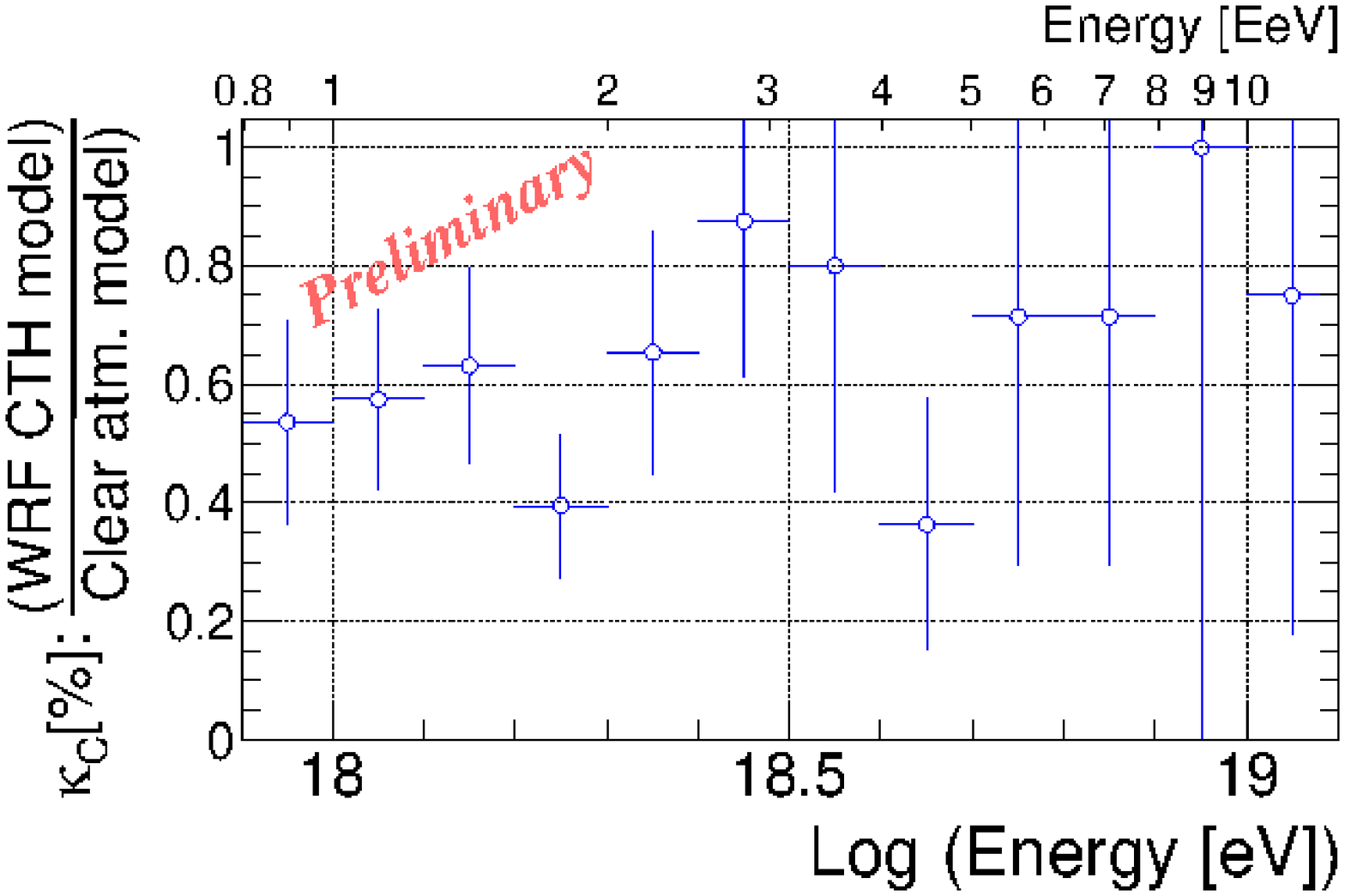} 
  \caption{Cloud efficiency as a function of the energy. Error bars denote the
    uncertainty due to simulated statistics with the UHECR spectrum. }
  \label{kappa}}&
    \parbox{0.45\textwidth}{      
      \includegraphics[width=0.45\textwidth]{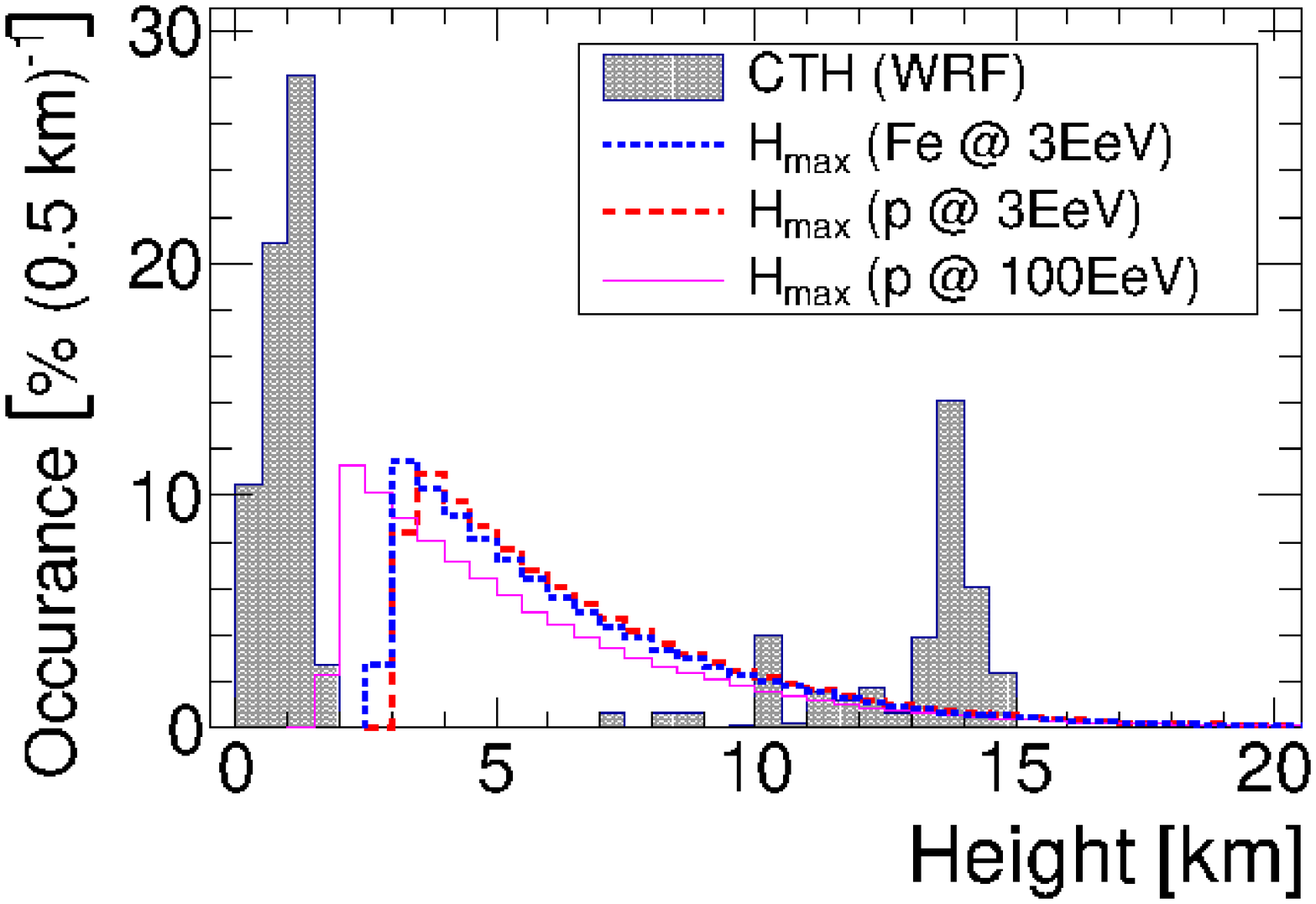} 
      \caption{CTH distribution from the WRF output during DAQ
        compared with those of the maximum height of air shower
        developments.}
      \label{cthhist32}}
  \end{tabular}
\end{figure*}

Figure~\ref{kappa} displays the cloud efficiency as a function of the primary
energy. Error bars denote the uncertainty due to simulated statistics with the
UHECR spectrum.

For this spectrum, the overall $\langle\kappa\rangle$ value is $\sim$62\%
above 1~EeV. 
We used a conservative threshold of the cloud fraction of 0.2 from the WRF
outputs that tends to give a CTH above the height that effectively
matters to the photon propagation from the air showers. One-layer optically
thick model clouds attenuate more than the usual case. The average background
level that includes the contribution from the clouds is also used in
simulations for the clear atmospheres. These assumed cloud properties allow
for setting a lower limit on the estimated number of detectable air shower
events for EUSO-SPB1.

Figure~\ref{cthhist32} displays the occurrence of the CTH from the WRF output
during the DAQ in the air shower detection mode. The maximum height of
shower developments are compared by assuming the typical $X_{\rm max}$ values
of $\sim$780~g~cm$^{-2}$ and $\sim$680~g~cm$^{-2}$ for$\sim$3~EeV proton and
iron UHECRs, respectively from Reference~\cite{Bellido19}. A $\sim$100~EeV
proton case of $X_{\rm max}=860$~g~cm$^{-2}$ that is relevant for the
full-scale missions is also shown. 

Two distinct populations are found. One is below ~1.5~km and the other is
above 7~km. The lower CTH
population accounts for the $\sim$60\% occurrence. In such situations, a
substantial portion of fluorescence photons from the air showers originate
above CTH is compared in Figure~\ref{hflux} and are not affected towards
EUSO-SPB1. The higher CTH population practically prevents the
photons in the 
other way round. The $\langle\kappa\rangle$ value of $\sim$60\% is naturally
explained by this population and does not seem to depend much on primary
particles in the EeV energies. 

\section{Summary and outlook}

From the present work, more than $\sim$60\% of the air showers in the EeV
regime that would trigger EUSO-SPB1 in clear atmosphere conditions also 
trigger even in the cloudy conditions based on the WRF output. 
The expected number of the detected air showers during the DAQ time of
EUSO-SPB1 is still on the order of one or fewer. Considering SPB's $\sim$100
day long flight potential, high probability of detecting shower events was
expected if flight could have been longer. Now, the collaboration is in preparation of the `EUSO-SPB2' mission anticipating launch in 2022.

In the full-scale space-based missions like K-EUSO and POEMMA, more than
$\sim$50~PDMs will be deployed to cover even wider FOV. They allow for
measurement of the entire
picture of the shower development above a few times 10~EeV energies.
The science cases require not only high statistics of UHECR observations but
also determination of exposures throughout continuously varying conditions.
EUSO-SPB1 was the first platform with the autonomous trigger to study such
situations. With a 25~cm telescope onboard the ISS, the Mini-EUSO mission will
be carried out in 2019 mainly aimed at measurement of the background. 
Experience, knowledge, and developed methods gained through the pathfinder
missions will be applied in the future missions.

\section*{Acknowledgments} 
\noindent{\normalsize
This work was partially supported by NASA grants NNX13AH54G,
NNX13AH52G, French Space Agency~(CNES), Italian Space Agency through the ASI INFN
agreement n.~2017-8-H.0, Italian Ministry of Foreign Affairs and International Cooperation, the Basic
Science Interdisciplinary Research Projects of RIKEN and JSPS KAKENHI Grant~(22340063, 23340081,
and 24244042), Deutsches Zentrum f\"ur Luft und Raumfahrt, and `Helmholtz Alliance for
Astroparticle Physics HAP' funded by the Initiative and Networking Fund of Helmholtz Association~(Germany).
We acknowledge NASA Balloon Program Office, Columbia Scientific Balloon Facility, 
Telescope Array Collaboration, and Wanaka airport. The present research used resources
of the National Energy Research Scientific Computing Center~(NERSC), a US~Department of Energy
Office of Science User Facility operated under Contract No.~DE-AC02-05CH11231.

}
\end{document}